%% file: main.tex
\PassOptionsToPackage{table,xcdraw}{xcolor}
\documentclass[sigconf,nonacm]{acmart}

\makeatletter
\patchcmd{\@typeset@author@bx}
  {\@authorfont\@currentauthors}
  {\@authorfont{\def\par{\ifhmode,\space\fi}\@currentauthors}}
  {}{\PackageError{main}{Could not format the author byline}{Check the acmart author layout.}}
\makeatother

\AtBeginDocument{%
  }

\usepackage{graphicx}
\usepackage{textcomp}
\usepackage{fontawesome}
\usepackage{url}
\usepackage{xspace}
\usepackage{enumitem}
\usepackage{multirow}
\usepackage{tikz}
\usepackage{pifont}
\usepackage{ragged2e}

\usepackage{algorithm}
\usepackage{algpseudocode}

\usepackage{amsmath,amsthm}
\usepackage{bm}
\usepackage{bbm}

\usepackage{mathtools}
\usepackage{dsfont}

\definecolor{kwblue}{RGB}{0, 60, 140}
\definecolor{commentgreen}{RGB}{34, 120, 50}
\definecolor{funcpurple}{RGB}{120, 30, 120}
\definecolor{gold}{RGB}{255,215,0}    % Gold
\definecolor{silver}{RGB}{192,192,192} % Silver
\definecolor{bronze}{RGB}{205,127,50}  % Bronze
\definecolor{orange}{RGB}{255,220,147}  % Bronze
\definecolor{pink}{RGB}{255,215,201}  % Bronze
 
\algrenewcommand\algorithmicfunction{\textcolor{kwblue}{\textbf{Function}}}
\algrenewcommand\algorithmicif{\textcolor{kwblue}{\textbf{if}}}
\algrenewcommand\algorithmicthen{\textcolor{kwblue}{\textbf{then}}}
\algrenewcommand\algorithmicelse{\textcolor{kwblue}{\textbf{else}}}
\algrenewcommand\algorithmicfor{\textcolor{kwblue}{\textbf{for}}}
\algrenewcommand\algorithmicwhile{\textcolor{kwblue}{\textbf{while}}}
\algrenewcommand\algorithmicdo{\textcolor{kwblue}{\textbf{do}}}
\algrenewcommand\algorithmicend{\textcolor{kwblue}{\textbf{end}}}
\algrenewcommand\algorithmicreturn{\textcolor{kwblue}{\textbf{return}}}
\algrenewcommand\algorithmicrepeat{\textcolor{kwblue}{\textbf{repeat}}}
\algrenewcommand\algorithmicuntil{\textcolor{kwblue}{\textbf{until}}}
 
\newcommand{\func}[1]{\textsc{\textcolor{funcpurple}{#1}}}
\newcommand{\cmt}[1]{\textcolor{commentgreen}{\(\triangleright\) \textit{#1}}}

\newcommand{\cmark}{\ding{51}}
\newcommand{\xmark}{\ding{55}}

\newcommand{\sysname}{\textsc{AgentDRC}\xspace}
\newcommand{\ballnumber}[1]{%
  \tikz[baseline=(myanchor.base)]%
  \node[circle,fill=black,inner sep=1pt] (myanchor) {\color{white}\bfseries\footnotesize #1};%
}
\setcopyright{none}
\renewcommand\footnotetextcopyrightpermission[1]{}
\begin{document}

\title{\sysname: Multi-\underline{Agent} Algorithms for Spatially-Aware Automated \underline{DRC} Correction}

\author{Pengju Liu}
\author{Nuo Xu}
\author{Winson Chen}
\author{Peijun Li}
\author{Yang (Katie) Zhao}
\author{Yu Cao}
\author{Caiwen Ding}
\affiliation{%
  \institution{University of Minnesota Twin Cities}
  \city{Minneapolis}
  \state{Minnesota}
  \country{USA}}
\renewcommand{\shortauthors}{Liu et al.}

\keywords{DRC Fixing, Multi-agent system}


\input{Sections/00_abstract}

\maketitle
\input{Sections/01_introduction}
\input{Sections/02_related_work}
\input{Sections/03_methodology}

\input{Sections/04_evaluation}

\input{Sections/05_result}
\input{Sections/06_conclusion}

% \clearpage
\clearpage
\bibliographystyle{ACM-Reference-Format}

\bibliography{reference.bib}

\end{document}

%% file: Sections/00_abstract.tex
\begin{abstract}
Design Rule Checking (DRC) error fixing remains a time-consuming bottleneck in physical design, often requiring manual intervention despite advanced EDA optimizations. Existing LLM-agent approaches fall short due to redundant contextual shapes, limited spatial reasoning over raw coordinates, and inefficient flat layout representations. We propose \sysname, which formulates iterative DRC correction as a multi-agent algorithm and introduces a spatial-aware verification and correction algorithm that reconstructs structured spatial relations among critical shapes and validates fixes through geometric and semantic constraint checking. The system coordinates four specialized agents to perceive multi-level, multi-modal geometric features, filter irrelevant shapes, and iteratively correct violations. Experiments show that \sysname achieves up to 30.4\% pass@1 and 55.7\% pass@5 on challenging violations, with 13.8\% pass@5 improvement over the best baseline. We will open-source all code and benchmark tasks upon acceptance.
\keywords{DRC Fixing, Multi-agent system, LLM Agent, Spatial Reasoning}
\end{abstract}

%% file: Sections/01_introduction.tex
\section{Introduction}
Large language model (LLM)-based agentic systems have recently achieved impressive results in code generation \cite{hong2024metagpt, zhang-etal-2024-codeagent, chen2024teaching}, robotics \cite{yao2023react, palm}, and mathematical reasoning \cite{toolformer, pal}, by autonomously decomposing complex tasks, leveraging external tools \cite{patil2024gorilla}, and performing step-by-step reasoning. These capabilities---tool use, multi-step planning, and contextual understanding---naturally align with the demands of physical design tasks that require iterative interaction with layout environments.

In contrast, the use of LLM-based agents in electronic design automation (EDA) remains largely unexplored, especially for design rule check (DRC) error fixing. This task requires modifying layout geometries to satisfy process design kit (PDK) constraints~\cite{siemens2018pdk, clark2016asap7}.
Modern EDA tools, such as Cadence Innovus~\cite{cadence2023innovus} and Siemens Calibre~\cite{siemens2020drc}, can automatically resolve part of the violations. However, many DRC errors still require manual intervention, which can take weeks for large-scale designs~\cite{siemens2020sadp, siemens2023drcdebug}.
These limitations highlight the need for more effective automated DRC fixing solutions~\cite{gfmentor2019drcfix}.
 
\begin{figure}[tbp]
    \centering
    \includegraphics[width=\linewidth]{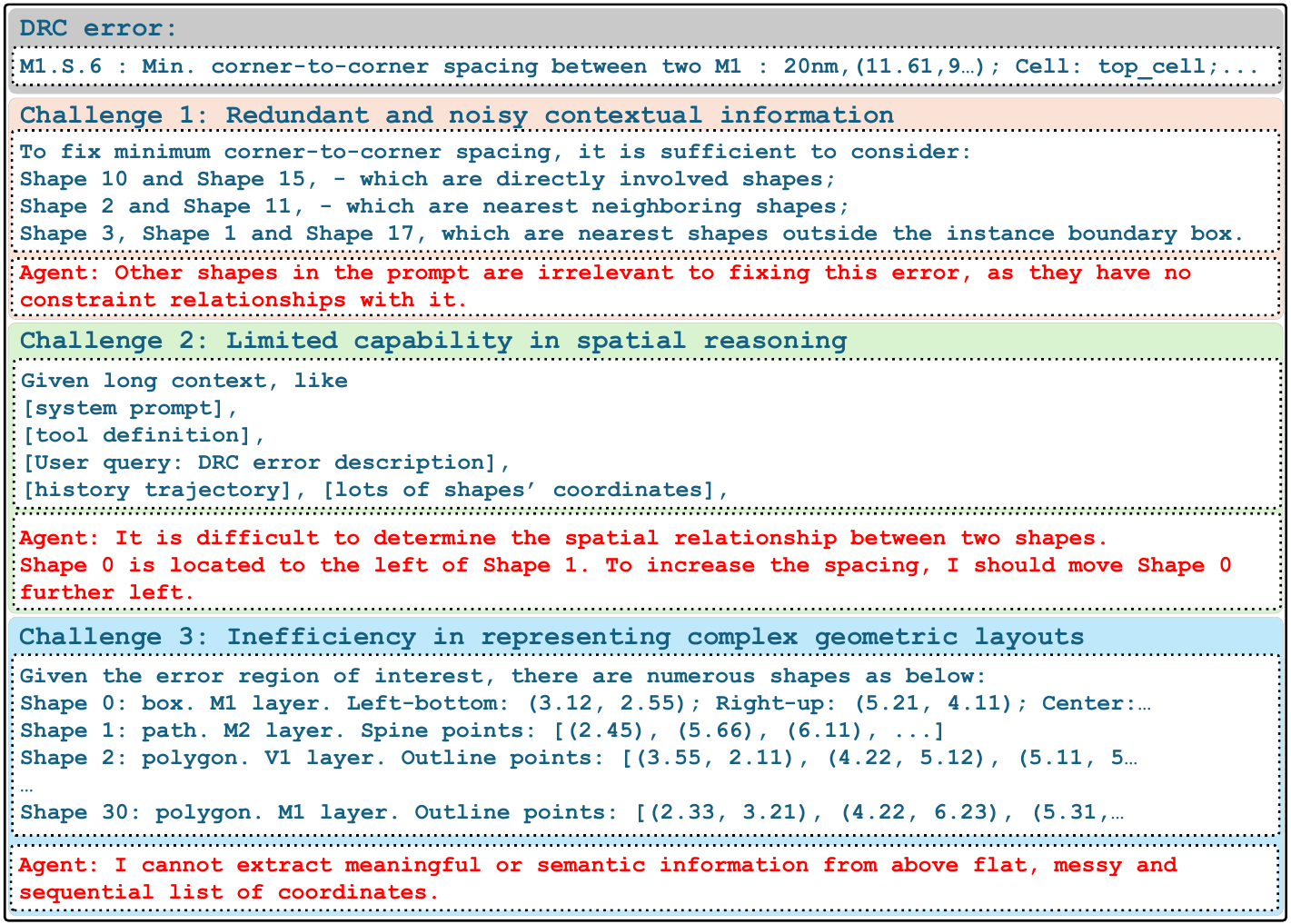}
    \caption{
Illustration of challenges in Agent-based DRC error understanding and resolution: (1) redundant surrounding shapes that distract the agent from constraint-related ones; (2) limited ability to infer spatial relationships and corrective actions from coordinate data; and (3) difficulty extracting meaningful geometric or semantic information from flat, sequential representations.
}
    \label{fig:intro}
\end{figure}
 
Motivated by this, we explore the potential of adopting LLM-based agents to automatically fix DRC errors. However, several challenges arise.
\textbf{Redundant and noisy contextual information (Challenge 1).}
The region around a DRC error often contains many nearby shapes across multiple layers, yet most of them share no constraints with the violation itself. Without a principled filtering mechanism, these unrelated shapes introduce significant noise and waste the agent's reasoning capacity, leading to incorrect or suboptimal fixes (see Fig.~\ref{fig:intro}).
\textbf{Limited capability in spatial reasoning (Challenge 2).}
Even when the relevant shapes have been identified, the agent struggles to interpret spatial relationships---relative direction, distance, and containment---from raw coordinate data, yet these are essential for determining corrective actions and detecting conflicts with nearby shapes (see Fig.~\ref{fig:intro}).
\textbf{Inefficiency in representing complex geometric layouts (Challenge 3).}
The flat, sequential coordinate data used to describe layout shapes fails to capture structural patterns or hierarchical context, creating a significant information bottleneck for high-level spatial reasoning required for DRC correction (Fig.~\ref{fig:intro}).
 
To address these challenges, we propose \textbf{\sysname}, which introduces a set of algorithms for spatially-aware DRC correction that coordinate multiple LLM agents to iteratively diagnose and fix violations through tool invocation and multi-turn interactions with the layout. To the best of our knowledge, \sysname is the first work to utilize LLM agents to fix DRC errors automatically. Our contributions are:
 
\begin{itemize}[leftmargin=*,label={$\triangleright$}]
 
\item We formalize DRC correction as an iterative multi-agent algorithm (Algorithm~1) that partitions the action space into geometric and non-geometric subsets and adaptively routes each proposed action through a spatial-aware geometric verification path or a semantic-only verification path, providing principled coordination for the multi-turn correction process.
 
\item To address \textbf{Challenge 1}, we design a Context Selection Agent that performs two-step filtering on multi-level shape descriptions: first identifying target shapes directly involved in the violation, then selecting up to $k$ constraint-relevant neighbors based on layer adjacency, spatial proximity, and constraint relevance. This effectively reduces the geometric context from dozens of shapes to a compact critical subset, significantly improving downstream spatial reasoning accuracy (\S\ref{sec:csa}).
 
\item To address \textbf{Challenge 2}, we design a \emph{spatial-aware verification and correction algorithm} (Algorithm~2) with two phases: a Spatial-Awareness Agent reconstructs structured spatial relations (distance, direction, containment) among critical shapes, transforming raw coordinates into interpretable representations; a Checker then verifies the action against geometric criteria---direction validity, magnitude sufficiency, and conflict detection---and auto-corrects deficiencies. For non-geometric actions, the algorithm further validates layer completeness and tool consistency via an additional semantic branch (\S\ref{sec:srac}).
 
\item To address \textbf{Challenge 3}, we design a multi-level, multi-modal observation mechanism that mimics how human engineers inspect layouts. The geometric context is decomposed into three hierarchical levels---directly involved, immediate context, and neighboring shapes---and presented in both textual and visual modalities when supported, enabling richer spatial perception at multiple granularities (\S\ref{sec:agent_impl}).
 
\end{itemize}
 
Comprehensive experiments demonstrate the effectiveness and generalizability of our proposed algorithms. On a challenging DRC benchmark, it achieves 13.8\%, 9.3\% and 8.6\% improvements in pass@5 compared with the best baseline based on GPT5-mini, Gemini-2.0 and DeepSeek-V3.1, respectively.

%% file: Sections/02_related_work.tex
\section{Related Work}

\noindent\textbf{Design Rule Check (DRC).}
Before tape-out, a circuit layout must comply with the physical design rules defined in the process design kit (PDK). Violating these rules can lead to fabrication failure or functional errors. For example, in the ASAP7 PDK~\cite{clark2016asap7}, the minimum width for metal layer M1 is 7\,nm; any M1 wire narrower than this violates manufacturability constraints. Hence, achieving a DRC-clean layout is essential for successful fabrication. However, even with advanced EDA tools such as Calibre \cite{mentor2016calibre}, Innovus \cite{cadence2023innovus}, or open-source OpenROAD \cite{openroad2020}, residual DRC violations remain and often require time-consuming manual correction.

\noindent\textbf{Single-Agent Systems.}
Agent-based systems have become a key paradigm for general-purpose intelligent behavior \cite{hong2024metagpt, zhang-etal-2024-codeagent, yao2023react, pal, palm, patil2024gorilla}. Modern AI models—such as OpenAI’s GPT, Google’s Gemini, and Anthropic’s Claude—employ agent-based architectures to enhance reasoning and tool use. In a single-agent system, one agent performs both reasoning and action within a unified loop, as exemplified by the ReAct framework \cite{yao2023react}.

\noindent\textbf{Multi-Agent Systems.}
For complex tasks, multi-agent systems offer greater scalability by assigning specialized roles to multiple agents working toward a shared goal \cite{zhang2021multiagent, bo2024reflective, nayak2024llamar}. A typical architecture adopts a \textit{planner–executor} scheme, where the planner decomposes the task and executors perform concrete actions \cite{wang2024voyager, song2023llmplanner}. This structure improves coordination, reasoning depth, and task efficiency across diverse domains.

\begin{figure*}[tbp]  % 图形浮动体，htbp表示在这里、顶部、底部或单独的浮动页
    \centering  % 使图形居中
    \includegraphics[width=1\linewidth]{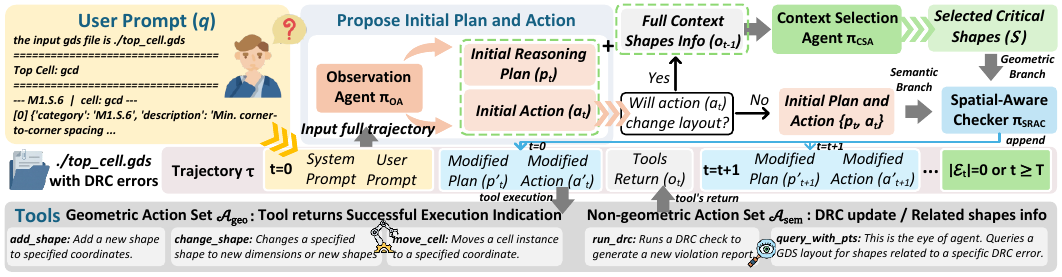}  % 插入图像，设置宽度为文本宽度的80%
    \caption{
\textbf{Overview of the proposed multi-agent algorithm, \sysname, for automatic DRC error fixing.}
}

    \label{fig:workflow}  % 图像标签，用于引用
\end{figure*}

%% file: Sections/03_methodology.tex
\section{Problem Formulation}

% We aim to design an intelligent agent system that automatically fixes design rule check (DRC) errors through iterative interactions with the target layout.

\subsection{Initial State $s_0$}
At the beginning, the user provides a system prompt $s$ and a user prompt $q$---containing the GDS layout $\mathcal{L}$ path and the initial set of known DRC violations $\mathcal{E}_0$---forming the initial state:
\[
s_0 = \{s,\, q\}.
\]
This corresponds to the scenario where an engineer performs placement and routing using conventional EDA tools, followed by several automatic DRC-fixing iterations. However, even after multiple runs, a number of DRC violations remain unresolved, which typically require manual correction by human engineers.

\subsection{Action Space $\mathcal{A}$}
% \textcolor{red}{(cd: where to use $\mathcal{A}$ in the text body?)}
The agent interacts with the layout environment through a discrete set of tool-based actions:
\begin{itemize}[leftmargin=*,label={$\triangleright$}]
\item \texttt{query\_with\_pts}: uses the coordinates of error markers provided in the DRC report to access related shapes. The agent can also specify the layers to inspect (e.g., M1, V1, etc.).
\item \texttt{move\_cell}: moves a specific cell instance by setting its new $(x, y)$ coordinates.
\item \texttt{change\_shape}: modifies the geometry of a specified shape, either within a specific cell instance or across all instances.
\item \texttt{add\_shape}: adds a new shape on a specific layer at the designated location.
\item \texttt{run\_drc}: executes the DRC engine to validate whether the current layout state is DRC-clean or if new violations are introduced.
\end{itemize}

% \subsection{Observation $\mathcal{O}$}
% At each step $t$, the agent receives an observation $o_t \in \mathcal{O}$: 
% % \textcolor{red}{cd: define $\mathcal{O}}
% \[
% o_t =
% \begin{cases}
% s_0, & t=0 \text{ (user prompt)},\\[3pt]
% f(a_{t-1}, s_{t-1}), & t>0 \text{ (tool feedback)}.
% \end{cases}
% \]
% % Observation provides feedback on the success of layout operations, updated DRC reports, or the spatial context around the error regions. Here, $f(\cdot)$ is the tool-feedback function that returns DRC reports, success/failure messages, or spatial information related to queried shapes; $s_{t-1}$ is the previous environment state and the environment state is represented as $s_t = \{o_t, a_{t-1}\}$; and $a_{t-1}$ is the action executed by the agent.
%  % \textcolor{red}{(CD: define f(), a, s.}
% Observation provides feedback on layout operations, DRC results, and the spatial context surrounding error regions. Here, $f(\cdot)$ denotes the tool-feedback function that generates this information based on the previous environment state $s_{t-1}$ and the action executed by the agent, $a_{t-1}$. The environment state at iteration $t$ is defined as $s_t = \{o_t, a_{t-1}\}$.
\subsection{Observation $\mathcal{O}$}

At each step $t$, the agent receives an observation $o_t \in \mathcal{O}$, where $\mathcal{O}$ is the space of all observable signals returned by the environment, including user prompts, tool feedback, updated DRC reports, and geometric context:

\[
o_t =
\begin{cases}
s_0, & t=0 \text{ (initial prompt)},\\[3pt]
\mathcal{T}(a_{t-1},\, \mathcal{L}), & t>0 \text{ (tool feedback)}.
\end{cases}
\]
Here, $\mathcal{T}(\cdot)$ is the environment transition function that executes an action on the current layout $\mathcal{L}$ and returns the resulting observation (e.g., queried shapes, operation feedback, or an updated DRC report). The agent maintains a full interaction trajectory $\tau$, initialized as $\tau \gets s_0$ and updated at each step as $\tau \gets \tau \cup \{p_t,\, a_t,\, o_t\}$, where $p_t$ is the reasoning plan and $a_t$ is the action taken at step $t$.

\subsection{Objective}
% At iteration $t$, the environment state is represented as $s_t = \{o_t, a_{t-1}\}$.  
The agent system $\pi_\theta$ determines at each step a reasoning plan $p_t$ and candidate action $a_t$ based on the accumulated trajectory:
% \[
$(p_t,\, a_t) = \pi_\theta(\tau)$,
% \]
and the environment evolves via the transition function:
% \[
$o_t = \mathcal{T}(a_t,\, \mathcal{L})$.
% \]
The objective is to minimize the number of remaining DRC violations after $T$ iterations:
% \[
$\min_{\pi_\theta} |\mathcal{E}_T|$.
% \]
Here, $\pi_\theta$ abstracts the full multi-agent architecture, and $\mathcal{E}_T$ is the final DRC violation set returned after $T$ iterations.

\section{Multi-agent algorithm for fixing DRC errors}
In this work, we propose a multi-agent algorithm, shown in Fig.~\ref{fig:workflow}. Starting from a user prompt and a GDS layout, the system performs iterative reasoning and action planning through cooperative agents.
\subsection{Overview}
\label{sec:overview}
We formalize the \textsc{AgentDRC} correction procedure in Algorithm~\ref{alg:main}. 
The algorithm takes as input a system prompt $s$, a user prompt $q$---containing the GDS file path and a known set of DRC violations $\mathcal{E}_0$---along with a maximum iteration budget $T$, and outputs a modified layout $\mathcal{L}^*$ with minimized violations. Firstly, we partition the action space $\mathcal{A}$ into two disjoint subsets:
\begin{equation}
  \mathcal{A} = \underbrace{\mathcal{A}_{\text{geo}}}_{\texttt{move\_cell},\, \texttt{change\_shape},\, \texttt{add\_shape}} \;\cup\; \underbrace{\mathcal{A}_{\text{sem}}}_{\texttt{query\_with\_pts},\, \texttt{run\_drc}},
\end{equation}
which drives an adaptive branching mechanism: geometric actions, which directly modify layout geometry, trigger the full pipeline of context selection (\S\ref{sec:csa}) and spatial-aware verification (\S\ref{sec:srac}), while non-geometric actions, which query contextual information or validate the current violations set, are routed through semantic-only verification. At each iteration~$t$, the algorithm proceeds through four stages, as illustrated in Fig.~\ref{fig:workflow} and Algorithm~\ref{alg:main}:
 
\begin{itemize}[leftmargin=*]
  \item \textbf{Stage 1} (Lines 5--6): the Observation Agent $\pi_{\text{OA}}$ consumes the full interaction trajectory $\tau$ and produces an initial reasoning plan $p_t$ together with a candidate action $a_t \in \mathcal{A}_{\text{geo}} \cup \mathcal{A}_{\text{sem}}$.
A branching indicator $\mathbb{I}_{\text{geo}} = \mathbbm{1}[a_t \in \mathcal{A}_{\text{geo}}]$ then determines which verification path to follow.
  \item \textbf{Stage 2} (Lines 9--10): If $a_t \in \mathcal{A}_{\text{geo}}$, the Context Selection Agent $\pi_{\text{CSA}}$ (\S\ref{sec:csa}) filters the geometric context from the prior tool feedback $o_{t-1}$ (typically the multi-level shape descriptions returned by \texttt{query\_with\_pts}) to produce a compact critical shape set $\mathcal{S}$.
  \item \textbf{Stage 3} (Lines 11--15): The Spatial-Aware Checker $\pi_{\text{SRAC}}$ verifies and corrects $(p_t, a_t)$ using $\mathcal{S}$ (geometric branch) or $\emptyset$ (semantic branch), as detailed in Algorithm~\ref{alg:verify}.
  \item \textbf{Stage 4} (Lines 17--20): The verified action $a_t'$ is executed via the environment transition function $\mathcal{T}$, which returns an observation $o_t$ (e.g., tool feedback, updated shapes, or a new DRC report). The trajectory is updated as $\tau \gets \tau \cup \{p_t', a_t', o_t\}$, and the violation set $\mathcal{E}_t$ is refreshed whenever $a_t' = \texttt{run\_drc}$.
\end{itemize}
 % ============================================================
% Algorithm 1: Main AgentDRC Loop
% ============================================================
\begin{algorithm}[t]
\caption{\textsc{AgentDRC}: Iterative Multi-Agent DRC Correction}
\label{alg:main}
\small
\begin{algorithmic}[1]
\Require System prompt $s$, user prompt $q$ (containing GDS path and known DRC violations $\mathcal{E}_0$), max iterations $T$
\Ensure Modified layout $\mathcal{L}^*$ with minimized DRC violations
\State $\tau \gets \{s, q\}$;\; $t \gets 0$
\State Define $\mathcal{A}_{\text{geo}} \triangleq \{\texttt{move\_cell},\, \texttt{change\_shape},\, \texttt{add\_shape}\}$ \Comment{\cmt{Geometric action set}}
\State Define $\mathcal{A}_{\text{sem}} \triangleq \{\texttt{query\_with\_pts},\, \texttt{run\_drc}\}$ \Comment{\cmt{Non-geometric action set}}
\Repeat
    \State \cmt{Stage 1: Observation Agent}
    \State $(p_t,\, a_t) \gets \pi_{\text{OA}}(\tau)$
    \Comment{\cmt{Generate reasoning $p_t$ and candidate action $a_t \in \mathcal{A}_{\text{geo}} \cup \mathcal{A}_{\text{sem}}$}}
    \State \cmt{Stage 2: Context Selection Agent ($\pi_{\text{CSA}}$)}
    \State $\mathbb{I}_{\text{geo}} \gets \mathbbm{1}[a_t \in \mathcal{A}_{\text{geo}}]$
    \Comment{\cmt{Branching indicator}}
    \If{$\mathbb{I}_{\text{geo}} = 1$}
        \State $\mathcal{S} \gets \pi_{\text{CSA}}(o_{t-1},\, p_t,\, a_t)$
        \Comment{\cmt{\S\ref{sec:csa}; select critical shapes from prior tool feedback}}
        \State \cmt{Stage 3: Spatial-Aware Checker ($\pi_{\text{SRAC}}$) --- geometric branch}
        \State $(p_t',\, a_t') \gets \pi_{\text{SRAC}}(p_t,\, a_t,\, \mathcal{S})$
        \Comment{\cmt{Alg.~\ref{alg:verify}; spatial agent + checker}}
    \Else
        \State \cmt{Stage 3: Spatial-Aware Checker ($\pi_{\text{SRAC}}$) --- semantic branch}
        \State $(p_t',\, a_t') \gets \pi_{\text{SRAC}}(p_t,\, a_t,\, \emptyset)$
        \Comment{\cmt{Alg.~\ref{alg:verify}; semantic-only verification \& correction}}
    \EndIf
    \State \cmt{Stage 4: Tool execution and state transition}
    \State $o_t \gets \mathcal{T}(a_t',\, \mathcal{L})$
    \Comment{\cmt{Execute action; environment returns observation $o_t$}}
    \State $\tau \gets \tau \cup \{p_t',\, a_t',\, o_t\}$;\; $t \gets t + 1$
    \State $\mathcal{E}_t \gets \mathcal{E}_t|_{o_t}$
    \Comment{\cmt{Violation set updated when $a_t' = \texttt{run\_drc}$}}
\Until{$|\mathcal{E}_t| = 0$ \textbf{or} $t \geq T$}
\State \Return $\mathcal{L}^* \gets \arg\min_{\mathcal{L} \in \{s_0,\ldots,s_t\}} |\mathcal{E}(\mathcal{L})|$
\end{algorithmic}
\end{algorithm}
The loop terminates when all violations are resolved ($|\mathcal{E}_t| = 0$) or the iteration budget $T$ is exhausted, returning the layout with the fewest remaining violations encountered during the trajectory.

\subsection{Context Selection Agent}
\label{sec:csa}

A key challenge in DRC fixing is that the geometric context around the target DRC marker often contains dozens of shapes across multiple layers, most of which are irrelevant to the violation at hand. Feeding all of them into downstream reasoning introduces significant noise and degrades the agent's spatial reasoning accuracy.

Given the tool feedback $o_{t-1}$ from \texttt{query\_with\_pts} containing multi-level shapes $\mathcal{C} = \{s_1, \ldots, s_N\}$ organized into Levels~1--3 (\S\ref{sec:agent_impl}), the current plan $p_t$, the proposed action $a_t$, and a neighbor budget $k$, the CSA (Context Selection Agent) produces a compact critical shape subset $\mathcal{S} \subseteq \mathcal{C}$. The selection proceeds in two steps designed to mimic the reasoning process of human engineers who first identify the shapes directly causing the violation, then examine only a small number of nearby shapes that may impose additional constraints.

\textbf{Step 1: Target Identification.}
The agent receives a structured prompt
\[
\mathcal{P}_1 \gets \langle\, \mathcal{C}^{(\ell_1)},\; p_t,\; a_t,\;\text{violation description}\,\rangle,
\]
containing the Level~1 shapes $\mathcal{C}^{(\ell_1)}$ (those that physically contact the DRC error marker). From this, $\pi_{\text{CSA}}$ identifies the set of target shapes directly involved in the reported violation:
\[
\mathcal{S}_{\text{target}} \gets \pi_{\text{CSA}}(\mathcal{P}_1).
\]
For example, for a minimum spacing error between two M1 edges, the agent selects the two M1 shapes whose edges are flagged by the DRC marker.

\textbf{Step 2: Constraint-Aware Neighbor Selection.}
The agent then receives a second prompt built from the Level~2--3 shapes together with the identified targets:
\[
\mathcal{P}_2 \gets \langle\, \mathcal{C}^{(\ell_2)} \cup \mathcal{C}^{(\ell_3)},\; \mathcal{S}_{\text{target}},\; p_t,\; a_t \,\rangle,
\]
augmented with three selection criteria: (i)~\emph{layer adjacency}---prioritizing shapes on neighboring layers (e.g., when the violation is on M1, also consider shapes on V0, V1, and M2); (ii)~\emph{spatial proximity}---favoring shapes nearest to the target shapes; and (iii)~\emph{constraint relevance}---selecting shapes that impose spacing, enclosure, or overlap rules related to the planned modification. The agent selects up to $k$ neighbor shapes most likely to affect the correctness of the proposed fix:
\[
\mathcal{S}_{\text{nbr}} \gets \pi_{\text{CSA}}(\mathcal{P}_2).
\]

The final output is the union of both sets,
\[
\mathcal{S} = \mathcal{S}_{\text{target}} \cup \mathcal{S}_{\text{nbr}},
\]
which is passed to the Spatial-Aware Checker. By reducing the context from $N$ shapes to a compact subset of typically $k + |\mathcal{S}_{\text{target}}|$ shapes (we set $k = 4$ in our experiments), the CSA significantly reduces contextual noise while preserving the information critical for constraint-aware reasoning.
% ============================================================
% Algorithm 3: Spatial-Aware Checker
% ============================================================
\begin{algorithm}[bt]
\caption{Spatial-Aware Checker ($\pi_{\text{SRAC}}$): Verification and Correction}
\label{alg:verify}
\small
\begin{algorithmic}[1]
\Require Critical shapes $\mathcal{S}$ (or $\emptyset$), plan $p_t$, action $a_t$
\Ensure Verified (and possibly corrected) plan and action $(p_t',\, a_t')$
\If{$\mathcal{S} = \emptyset$} \Comment{\cmt{Semantic Branch (non-geometric actions)}}
    \State $v_{\text{layer}} \gets \func{CheckLayerCompleteness}(a_t)$ \Comment{\cmt{All neighboring layers queried?}}
    \State $v_{\text{tool}} \gets \func{CheckToolConsistency}(a_t)$ \Comment{\cmt{Correct GDS file \& params?}}
    \If{$v_{\text{layer}} \wedge v_{\text{tool}}$}\; \Return $(p_t,\, a_t)$ \Comment{\cmt{Pass through unchanged}}
    \Else\; \Return $\func{Correct}(p_t,\, a_t,\, \text{diagnostics})$ \Comment{\cmt{Auto-correct and return}}
    \EndIf
\Else \Comment{\cmt{Geometric Branch (layout modification actions)}}
    \State \cmt{Phase I: Spatial Relation Construction (Spatial-Awareness Agent)}
    \State $\mathcal{R} \gets \emptyset$
\For{each pair $(s_i, s_j) \in \mathcal{S} \times \mathcal{S},\; i < j$}
    \State $d_{ij} \gets \func{MinDist}(\func{BBox}(s_i),\, \func{BBox}(s_j))$
    \State $\textit{dir}_{ij} \gets \func{RelDir}(\func{BBox}(s_i),\, \func{BBox}(s_j))$
    \State $\textit{cont}_{ij} \gets \func{Contain}(\func{BBox}(s_i),\, \func{BBox}(s_j))$
    % \addtocounter{ALG@line}{-1}
    \State $\mathcal{R} \gets \mathcal{R} \cup \{(s_i,\, s_j,\, (d_{ij},\, \textit{dir}_{ij},\, \textit{cont}_{ij}))\}$
\EndFor
    \State \cmt{Phase II: Constraint Verification (Checker) on proposed action}
    \State $v_{\text{dir}} \gets \func{ValDir}(a_t,\, \mathcal{R})$ \Comment{\cmt{Move direction resolves violation?}}
    \State $v_{\text{mag}} \gets \func{ValMag}(a_t,\, \mathcal{R})$ \Comment{\cmt{Sufficient displacement?}}
    \State $v_{\text{cfl}} \gets \func{DetConflict}(a_t,\, \mathcal{R})$ \Comment{\cmt{New violations introduced?}}
    \If{$v_{\text{dir}} \wedge v_{\text{mag}} \wedge \neg\, v_{\text{cfl}}$}\; \Return $(p_t,\, a_t)$
    \Else\; \Return $\func{Correct}(p_t,\, a_t,\, \mathcal{R},\, \text{diagnostics})$
    \EndIf
\EndIf
\end{algorithmic}
\end{algorithm}
\subsection{Spatial-Aware Verification and Correction}
\label{sec:srac}

Even after context selection, the proposed plan and action may still contain spatial reasoning errors---e.g., moving a shape in the wrong direction, displacing it by an insufficient amount to satisfy the spacing rule, inadvertently introducing new violations with neighboring shapes, or querying an incomplete set of layers and missing critical constraints. Algorithm~\ref{alg:verify} addresses this via a two-branch mechanism that verifies and, if necessary, corrects $(p_t, a_t)$ to produce $(p_t', a_t')$.
\textbf{Semantic Branch} ($\mathcal{S} = \emptyset$, Lines 2--6). For non-geometric actions (e.g., \texttt{query\_with\_pts}, \texttt{run\_drc}), $\pi_{\text{SRAC}}$ checks two criteria:
\begin{itemize}[leftmargin=*,label={$\triangleright$}]
  \item $v_{\text{layer}}$: queried layers cover all relevant neighbors (e.g., M1 violation $\Rightarrow$ also inspect V0, V1, M2);
  \item $v_{\text{tool}}$: tool calls reference the most recently modified GDS file rather than the original one, and all parameters (coordinates, instance indices, layer specifications) are correctly formatted and within valid ranges.
\end{itemize}
If $v_{\text{layer}} \wedge v_{\text{tool}}$, the pair passes through unchanged; otherwise it is auto-corrected.

\textbf{Geometric Branch} ($\mathcal{S} \neq \emptyset$, Lines 7--16). When the action modifies layout geometry, the checker proceeds in two phases.

\begin{figure}[t]
    \centering
    \includegraphics[width=\linewidth]{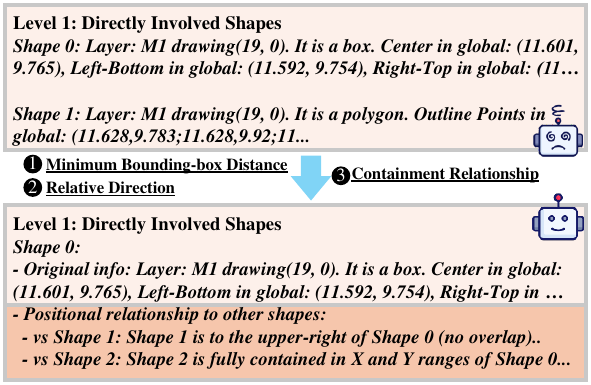}
    \caption{
\textbf{Structured Spatial Relations.}
The Spatial-Awareness Agent transforms raw geometric coordinates around a DRC error
into structured spatial relations.
% by
% (1) listing all shapes' boundary box,
% (2) identifying target shapes involved in the modification, and
% (3) inferring about their positional relationships
% (e.g., containment, adjacency, or direction).
% The bottom example shows the final summarized description received by the checker,
% which embeds spatial relations such as
% ``Shape~1 is to the upper-right of Shape~0'' and
% ``Shape~2 is contained within Shape~0.''
% This enables spatial-aware validation of layout modifications.
}

    \label{fig:spatial_aware}
\end{figure}

\emph{Phase I --- Spatial Relation Construction.}
A Spatial-Awareness Agent processes $\mathcal{S}$ and builds a structured relation set
\[
\mathcal{R} = \bigl\{(s_i,\, s_j,\, (d_{ij},\, \textit{dir}_{ij},\, \textit{cont}_{ij})) \;\big|\; s_i, s_j \in \mathcal{S},\; i < j\bigr\},
\]
where $d_{ij} = \textsc{MinDist}(\textsc{BBox}(s_i), \textsc{BBox}(s_j))$ is \ballnumber{1} the minimum bounding-box distance, $\textit{dir}_{ij}$ is \ballnumber{2} the relative direction (e.g., ``Shape~1 is to the upper-right of Shape~0''), and $\textit{cont}_{ij}$ captures \ballnumber{3} the containment relationship (full, partial, or none). This transforms raw coordinates into semantically meaningful spatial cues, as illustrated in Fig.~\ref{fig:spatial_aware}.

\emph{Phase II --- Constraint Verification.}
Given $\mathcal{R}$, the Checker evaluates three geometric criteria on $a_t$:
\begin{itemize}[leftmargin=*,label={$\triangleright$}]
  \item $v_{\text{dir}}$: the move direction resolves the violation (not toward the conflicting edge);
  \item $v_{\text{mag}}$: the displacement satisfies the minimum spacing rule;
  \item $v_{\text{cfl}}$: no new violations are introduced with neighboring shapes in $\mathcal{S}$.
\end{itemize}
If $v_{\text{dir}} \wedge v_{\text{mag}} \wedge \neg\, v_{\text{cfl}}$, the original pair is returned; otherwise $\pi_{\text{SRAC}}$ issues a corrected $(p_t', a_t')$ using $\mathcal{R}$ and diagnostic information.

By dynamically switching between the semantic and geometric branches, the algorithm ensures that every decision---whether a context query or a physical layout modification---is both logically consistent and spatially aware, maintaining correctness throughout the multi-turn correction process.
\subsection{Agent Implementation}
\label{sec:agent_impl}
 
This section describes the implementation details of the agents used in Algorithms~\ref{alg:main}--\ref{alg:verify}.
 
\textbf{Observation Agent ($\pi_{\text{OA}}$).}
The OA is the entry point of each iteration, perceiving the current state and proposing an initial plan $p_t$ and action $a_t$. Its inputs differ by iteration:
\begin{itemize}[leftmargin=*]
  \item \emph{$t=0$}: the system prompt and user prompt, containing the GDS path and known DRC violations $\mathcal{E}_0$.
  \item \emph{$t>0$}: the full accumulated trajectory $\tau$, which contains all prior plans, actions, and tool observations $o_{t'}$---including geometric context descriptions returned by \texttt{query\_with\_pts}, operation success indicators (e.g., whether a cell instance was successfully moved), and updated DRC reports from \texttt{run\_drc}.
\end{itemize}

\begin{figure}[tbp]
    \centering
    \includegraphics[width=0.9\linewidth]{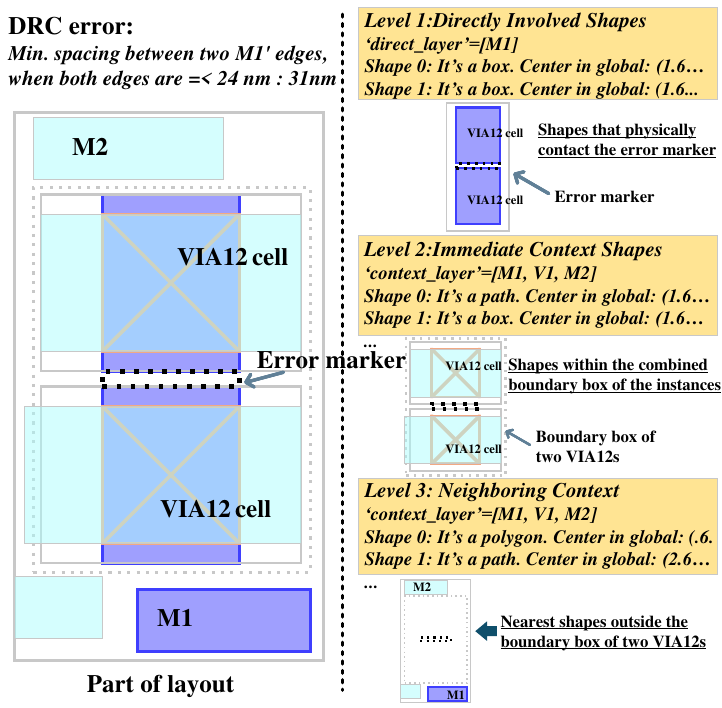}
    \caption{\textbf{Multi-Level and Multi-Modal Representation of Context Shapes.}
    (Left) Example of a DRC error and its affected layout region.
    (Right) Three context levels are constructed around the error marker.
    % \underline{Level~1} — Directly involved shapes that physically contact the error marker;
    % \underline{Level~2} — Shapes within the combined boundary box of the cell instances containing the Level~1 shapes;
    % \underline{Level~3} — Neighboring shapes located outside the combined boundary box.
    % This representation captures both local and global spatial context for DRC correction.
    }
    \label{fig:multi_level}
\end{figure}

To perceive the complex spatial context around each DRC error, we design a \emph{multi-level, multi-modal representation} that mimics how human engineers visually inspect layouts. As illustrated in Fig.~\ref{fig:multi_level}, the geometric context from \texttt{query\_with\_pts} is decomposed into three hierarchical levels:
\begin{itemize}[leftmargin=*]
  \item \underline{Level~1}: Shapes physically contacting the DRC error marker---corresponding to the zoomed-in view when an engineer double-clicks an error.
  \item \underline{Level~2}: Shapes within the combined bounding box of the cell instances containing the Level~1 shapes---emulating the slightly zoomed-out view for nearby constraints.
  \item \underline{Level~3}: Shapes just outside the combined bounding box---crucial for evaluating potential inter-instance conflicts during cell movements.
\end{itemize}
The agent can dynamically adjust the observation range (e.g., from the nearest to the $m$-th nearest shape), enabling adaptive compression of the geometric context. Both textual descriptions and visual renderings are provided when the base model supports multi-modal input (e.g., GPT-5-mini and Gemini-2.0), enabling richer spatial understanding.

Based on this multi-level, multi-modal context and the full trajectory $\tau$, the OA outputs an initial reasoning plan $p_t$ (e.g., ``To fix the spacing violation, I will use \texttt{move\_cell} to shift the instance further from the conflicting edge'') and a corresponding candidate action $a_t$ (e.g., \texttt{move\_cell(new\_x: 6.09, new\_y: 7.1)}).
 
\textbf{Context Selection Agent ($\pi_{\text{CSA}}$).}
As described in \S\ref{sec:csa}, the CSA is instantiated as an LLM agent that receives structured prompts containing the multi-level textual context and selection criteria. In our experiments, we set the neighbor budget to $k=4$.

\textbf{Spatial-Awareness Agent and Checker.}
Both are implemented as components of the Spatial-Aware Checker (Algorithm~\ref{alg:verify}). The Spatial-Awareness Agent receives the raw critical shapes selected by the CSA and outputs structured spatial relation descriptions in natural language. The Checker agent then combines these spatial relations with the proposed plan and action to produce its final verification and correction output.

%% file: Sections/04_evaluation.tex
\section{Evaluation}
% \usepackage{multirow}
\begin{table*}[]
\caption{Overall comparison on L1. L1 evaluates the base model’s fundamental abilities.}
\label{tab:l1-eval}
\begin{tabular}{lcccccccccc}
\hline
\multicolumn{1}{c|}{}                                             & \multicolumn{1}{c|}{}                              & \multicolumn{1}{c|}{}                              & \multicolumn{8}{c}{\textbf{L1}}                                                                                                                 \\ \cline{6-9}
\multicolumn{1}{c|}{}                                             & \multicolumn{1}{c|}{}                              & \multicolumn{1}{c|}{}                              & \multicolumn{2}{c}{class 1(\%)}    & \multicolumn{2}{c}{class 2(\%)}     & \multicolumn{2}{c}{class 3(\%)}      & \multicolumn{2}{c}{overall(\%)} \\
\multicolumn{1}{c|}{\multirow{-3}{*}{Method}}                     & \multicolumn{1}{c|}{\multirow{-3}{*}{Open Source}} & \multicolumn{1}{c|}{\multirow{-3}{*}{Model Size}} & pass@1 & pass@5                    & pass@1 & pass@5                    & pass@1 & pass@5                     & pass@1         & pass@5         \\ \hline
\multicolumn{11}{c}{\cellcolor[HTML]{C5C5C5}Different Base Models}                                                                                                                                                                                                       \\ \hline
\multicolumn{1}{l|}{\cellcolor[HTML]{FFDF93}ReAct(GPT5-mini)}           & \multicolumn{1}{c|}{\xmark}                             & \multicolumn{1}{c|}{N/A$^\dagger$}  & \cellcolor[HTML]{FFD800}94.8   & \multicolumn{1}{c|}{\cellcolor[HTML]{FFD800}98.9} & \cellcolor[HTML]{FFD800}91.7   & \multicolumn{1}{c|}{\cellcolor[HTML]{FFD800}99.9} & \cellcolor[HTML]{FFD800}96.7   & \multicolumn{1}{c|}{\cellcolor[HTML]{FFD800}100.0} & \cellcolor[HTML]{FFD800}94.3           & \cellcolor[HTML]{FFD800}99.2           \\
\multicolumn{1}{l|}{\cellcolor[HTML]{FFDF93}ReAct(Gemini-2.0)}        & \multicolumn{1}{c|}{\xmark}                             & \multicolumn{1}{c|}{N/A$^\dagger$}  & \cellcolor[HTML]{CD7F32}72.4   & \multicolumn{1}{c|}{\cellcolor[HTML]{C0C0C0}94.2} & \cellcolor[HTML]{CD7F32}53.3   & \multicolumn{1}{c|}{\cellcolor[HTML]{C0C0C0}95.8} & \cellcolor[HTML]{C0C0C0}90.0   & \multicolumn{1}{c|}{\cellcolor[HTML]{FFD800}100.0} & \cellcolor[HTML]{CD7F32}70.3           & \cellcolor[HTML]{C0C0C0}95.1           \\
\multicolumn{1}{l|}{\cellcolor[HTML]{FFDF93}ReAct(DeepSeek-V3.1)} & \multicolumn{1}{c|}{\cmark}                             & \multicolumn{1}{c|}{671B}           & \cellcolor[HTML]{C0C0C0}85.2   & \multicolumn{1}{c|}{\cellcolor[HTML]{C0C0C0}94.2} & \cellcolor[HTML]{C0C0C0}66.7   & \multicolumn{1}{c|}{\cellcolor[HTML]{CD7F32}91.3} & \cellcolor[HTML]{C0C0C0}90.0   & \multicolumn{1}{c|}{\cellcolor[HTML]{FFD800}100.0} & \cellcolor[HTML]{C0C0C0}82.0           & \cellcolor[HTML]{CD7F32}94.2           \\
\multicolumn{1}{l|}{\cellcolor[HTML]{FFDAC9}ReAct(Qwen-2.5-7B)}   & \multicolumn{1}{c|}{\cmark}                             & \multicolumn{1}{c|}{7B}             & 9.52   & \multicolumn{1}{c|}{26.7} & 6.67   & \multicolumn{1}{c|}{23.6} & 10.0   & \multicolumn{1}{c|}{42.6}  & 9.0            & 27.7           \\
\multicolumn{1}{l|}{\cellcolor[HTML]{FFDAC9}ReAct(Llama-3.1-8B)}  & \multicolumn{1}{c|}{\cmark}                             & \multicolumn{1}{c|}{8B}             & 2.38      & \multicolumn{1}{c|}{11.9}    & 0.00      & \multicolumn{1}{c|}{0.00}    & 13.3      & \multicolumn{1}{c|}{47.2}     & 3.00              & 13.1              \\ \hline
\multicolumn{11}{l}{\textcolor{orange}{\faStar} Orange rows denote base models strong enough to act as agents for DRC fixing.}                                                                                                                                                         \\
\multicolumn{11}{l}{\textcolor{pink}{\faBan} Pink rows denote base models with limited capacity for DRC fixing. $\dagger$: The model size of GPT5-mini and Gemini-2.0 is undisclosed.} 
\end{tabular}
\end{table*}
% \noindent\textbf{EDA Tool and Technology PDK.}
We use the open-source \texttt{OpenROAD} \cite{openroad2020} flow to generate layouts from Verilog designs and \texttt{KLayout} \cite{klayout2024} for layout operations. All experiments are conducted under the ASAP7  \cite{clark2016asap7}.

\subsection{Evaluation Setting}

\noindent\textbf{Benchmark.}
Since no public benchmark exists for DRC fixing, we construct a multi-level benchmark to evaluate agents under varying complexities. It contains two levels: \textbf{L1} for basic operations and \textbf{L2} for complex reasoning. \emph{Notably, the L2 tasks are collected from DRC violations that \texttt{OpenROAD} cannot resolve automatically, so comparing with \texttt{OpenROAD}'s built-in DRC fixer is unnecessary.
}

% \noindent\textbf{L1.}
\textbf{L1} evaluates the base model’s fundamental capabilities: tool invocation and basic DRC knowledge. All L1 errors are manually constructed. It comprises three classes:
\begin{itemize}[leftmargin=*, label=$\diamond$]
    \item \textbf{Class 1}: Atomic cases involving exactly one shape. For example, a single M1 rectangle with width 14\,nm violates the 18\,nm minimum width rule---the agent must widen it to at least 18\,nm.
    \item \textbf{Class 2}: Layouts with 2--3 shapes across different layers, extending Class~1 without requiring reasoning---the agent needs to handle multiple shapes when compared with Class 1. For example, an M1 rectangle that only partially overlaps a V1 shape below it must be enlarged or repositioned to achieve full enclosure.
    \item \textbf{Class 3}: Practical layouts with dense surrounding shapes, where a simple error has been injected by human engineers into a DRC-clean layout. The fix remains straightforward but the agent must handle complex context. For example, one M1 wire in a \texttt{gcd} placement is narrowed below the minimum width---the fix is simply to restore its width, but it absorbs amounts of shape in the dense context.
\end{itemize}

% \noindent\textbf{L2.}
\textbf{L2} evaluates the agent’s {reasoning ability} within complex contexts, {specifically targeting errors left unresolved by \texttt{Open\-ROAD} that typically necessitate human intervention}. We define three classes of increasing reasoning difficulty:
\begin{itemize}[leftmargin=*, label=$\diamond$]
    \item \textbf{Class 1}: Tasks requiring exactly one fix step, but where that step demands non-trivial geometric reasoning among surrounding shapes. For example, an \texttt{M1.S.6} corner-to-corner spacing violation (min. 20\,nm) cannot be resolved by moving the M1 shape alone---doing so would violate the V1 enclosure rule---so the agent must instead move the VIA12 cell.
    \item \textbf{Class 2}: Tasks requiring a minimum of multiple fix steps, where resolving the primary violation also disturbs neighboring shapes constrained by spacing or inter-layer rules. For example, fixing an \texttt{M1.S.2} edge-spacing violation (min. 25\,nm, triggered when one edge exceeds 36\,nm) requires moving the VIA12 together with its associated M1 and V1 shapes, and adding a new M1 shape to prevent newly introduced violations.
    \item \textbf{Class 3}: Tasks where the reported violation is not the root cause, so the agent cannot derive the minimum-step solution from the DRC report alone. For example, \texttt{V1.S.4} reports a min. corner-to-corner spacing of 27\,nm between two V1 instances---one with and one without a 5\,nm M2 end-cap. Directly increasing the V1 spacing would introduce new errors elsewhere, requiring even more steps. The minimum-step solution is instead to extend the M2 end-cap, which eliminates the asymmetric spacing rule that triggered the violation.
\end{itemize}
% \subsubsection{L1}
% L1 assesses the agent’s ability to understand DRC rules and correctly use EDA tools, focusing on tasks that require minimal reasoning.  
% It consists of three classes:
% \begin{itemize}[leftmargin=*]
%     \item \textbf{Class 1:} Single-shape layouts requiring direct solution as indicated by DRC reports.
%     \item \textbf{Class 2:} Simple layouts (fewer than three shapes) involving limited reasoning or cross-layer constraints. 
%     \item \textbf{Class 3:} Realistic layouts with many surrounding shapes but simple fixes, used to test context understanding.
% \end{itemize}

% \subsubsection{L2}
% L2 evaluates the agent’s reasoning ability on complex contexts using errors left unresolved by \texttt{OpenROAD}.  
% These tasks generally require human intervention in practice.  
% We define three representative classes:
% \begin{itemize}[leftmargin=*]
%     \item \textbf{Class 1:} Single-step reasoning on complex layouts.
%     \item \textbf{Class 2:} Multi-step fixes involving additional shapes constrained by spatial or inter-layer rules.
%     \item \textbf{Class 3:} Cases where the reported violation is not the root cause, requiring exploration of hidden constraints.
% \end{itemize}

\noindent\textbf{Metrics.}
Each task takes a system prompt $s$, a user prompt $q$ containing the GDS layout path and the initial violation set $\mathcal{E}_0$, and a maximum iteration budget $T$ as input, and outputs a modified layout $\mathcal{L}^*$. A solution is considered \emph{successful} if $|\mathcal{E}(\mathcal{L}^*)| = 0$ (i.e., fully DRC-clean), and a \emph{failure} otherwise---including cases where violations remain or new violations are introduced.
We report results using the \textit{pass@k} metric \cite{chen2021codex}, which measures the probability that at least one of $k$ independent runs produces a DRC-clean layout ($k \in \{1, 5\}$). We use $n=10$ samples per task at temperature 0.7.

\noindent\textbf{Baselines.}
We compare our method with the following frameworks:
% \begin{itemize}[leftmargin=*]
    % \item 
    \textit{ReAct (Single Agent):} Integrates reasoning and tool execution in a unified loop.
    % \item 
    \textit{Planner \& Actor (Multi-Agent):} A planner generates reasoning plans while an actor executes tools accordingly.
% \end{itemize}
We evaluate each architecture with three base models: GPT5-mini, Gemini-2.0, and DeepSeek-V3.1 (text modality-only).

%% file: Sections/05_result.tex
\subsection{Research Questions}
% \usepackage{multirow}
\begin{table*}[]
\caption{Comparison of \sysname against various baseline models on L2.}
\label{tab:main-r2}
\begin{tabular}{lcccccccccc}
\hline
\multicolumn{1}{c|}{}                         & \multicolumn{1}{c|}{}                              & \multicolumn{1}{c|}{}                           & \multicolumn{8}{c}{\textbf{L2}}                                                                                                                                                                     \\ \cline{6-9}
\multicolumn{1}{c|}{}                         & \multicolumn{1}{c|}{}                              & \multicolumn{1}{c|}{}                           & \multicolumn{2}{c}{class 1(\%)}     & \multicolumn{2}{c}{class 2(\%)}    & \multicolumn{2}{c}{class 3(\%)}                             & \multicolumn{2}{c}{overall(\%)}                             \\
\multicolumn{1}{c|}{\multirow{-3}{*}{Method}} & \multicolumn{1}{c|}{\multirow{-3}{*}{Open Source}} & \multicolumn{1}{c|}{\multirow{-3}{*}{Modality}} & pass@1 & pass@5                     & pass@1 & pass@5                    & pass@1 & pass@5                                            & pass@1                       & pass@5                       \\ \hline
\multicolumn{11}{c}{\cellcolor[HTML]{C5C5C5}Single-Agent}                                                                                                                                                                                                                                             \\ \hline
\multicolumn{1}{l|}{ReAct(GPT5-mini)}               & \multicolumn{1}{c|}{\xmark}  & \multicolumn{1}{c|}{T}                          & \cellcolor[HTML]{C0C0C0}24.0   & \multicolumn{1}{c|}{\cellcolor[HTML]{C0C0C0}51.7}  & \cellcolor[HTML]{C0C0C0}16.0   & \multicolumn{1}{c|}{34.8} & 6.7    & \multicolumn{1}{c|}{33.3}                         & \cellcolor[HTML]{C0C0C0}18.3 & \cellcolor[HTML]{CD7F32}41.9 \\
\multicolumn{1}{l|}{ReAct(Gemini-2.0)}            & \multicolumn{1}{c|}{\xmark}  & \multicolumn{1}{c|}{T}                          & 10.0   & \multicolumn{1}{c|}{29.1}  & 8.00   & \multicolumn{1}{c|}{27.5} & \cellcolor[HTML]{C0C0C0}16.7   & \multicolumn{1}{c|}{\cellcolor[HTML]{C0C0C0}56.5}                         & 10.0                         & 32.0                         \\
\multicolumn{1}{l|}{ReAct(DeepSeek-V3.1)}     & \multicolumn{1}{c|}{\cmark}  & \multicolumn{1}{c|}{T}                          & 10.0   & \multicolumn{1}{c|}{26.9}  & \cellcolor[HTML]{CD7F32}14.0   & \multicolumn{1}{c|}{37.5} & \cellcolor[HTML]{CD7F32}10.0   & \multicolumn{1}{c|}{\cellcolor[HTML]{CD7F32}50.0}                         & 11.7                         & 34.5                         \\ \hline
\multicolumn{11}{c}{\cellcolor[HTML]{C5C5C5}Multi-Agent System}                                                                                                                                                                                                                                       \\ \hline
\multicolumn{1}{l|}{MAS(GPT5-mini)}                 & \multicolumn{1}{c|}{\xmark}  & \multicolumn{1}{c|}{T}                          & 8.00   & \multicolumn{1}{c|}{27.5} & 6.00   & \multicolumn{1}{c|}{25.6} & 3.33   & \multicolumn{1}{c|}{16.7}                         & 6.52                         & 25.3                         \\
\multicolumn{1}{l|}{MAS(Gemini-2.0)}              & \multicolumn{1}{c|}{\xmark}  & \multicolumn{1}{c|}{T}                          & 13.0   & \multicolumn{1}{c|}{25.0}  & 8.00   & \multicolumn{1}{c|}{17.8} & 3.33   & \multicolumn{1}{c|}{16.7}                         & 9.56                         & 20.8                         \\
\multicolumn{1}{l|}{MAS(DeepSeek-V3.1)}            & \multicolumn{1}{c|}{\cmark}  & \multicolumn{1}{c|}{T}                          & 3.00   & \multicolumn{1}{c|}{15.0}  & 8.00   & \multicolumn{1}{c|}{\cellcolor[HTML]{CD7F32}37.8} & 6.67   & \multicolumn{1}{c|}{25.9}                         & 5.65                         & 26.3                         \\
\multicolumn{1}{l|}{\textbf{\sysname(GPT)}}              & \multicolumn{1}{c|}{\cmark}  & \multicolumn{1}{c|}{T+V}                        & \cellcolor[HTML]{FFD800}39.0   & \multicolumn{1}{c|}{\cellcolor[HTML]{FFD800}68.9}  & \cellcolor[HTML]{FFD800}22.0   & \multicolumn{1}{c|}{\cellcolor[HTML]{FFD800}42.7} & 6.67   & \multicolumn{1}{c|}{33.3}                         & \cellcolor[HTML]{FFD800}30.4 & \cellcolor[HTML]{FFD800}55.7 \\
\multicolumn{1}{l|}{\textbf{\sysname(Gemini)}}           & \multicolumn{1}{c|}{\cmark}  & \multicolumn{1}{c|}{T+V}                        & 17.0   & \multicolumn{1}{c|}{\cellcolor[HTML]{CD7F32}47.7}  & 7.00   & \multicolumn{1}{c|}{22.5} & \cellcolor[HTML]{FFD800}33.3   & \multicolumn{1}{c|}{\cellcolor[HTML]{FFD800}82.4} & 14.8                         & 41.3                         \\
\multicolumn{1}{l|}{\textbf{\sysname(DeepSeek)}}         & \multicolumn{1}{c|}{\cmark}  & \multicolumn{1}{c|}{T}                          & \cellcolor[HTML]{CD7F32}21.0   & \multicolumn{1}{c|}{44.5}  & \cellcolor[HTML]{C0C0C0}16.0   & \multicolumn{1}{c|}{\cellcolor[HTML]{C0C0C0}41.9} & \cellcolor[HTML]{CD7F32}10.0   & \multicolumn{1}{c|}{42.6}                         & \cellcolor[HTML]{CD7F32}17.4 & \cellcolor[HTML]{C0C0C0}43.1 \\ \hline
\multicolumn{11}{l}{$\dagger$: ranking of performance as \textcolor{gold}{\faTrophy} 1st, \textcolor{silver}{\faTrophy} 2nd, \textcolor{bronze}{\faTrophy} 3rd place; Note: MAS is Planner and Actor Architecture.}\\
\multicolumn{11}{l}{$\S$: “T” denotes multi-level text-only modality, while “T+V” indicates the use of both multi-level text and vision modalities.}    \\
\multicolumn{11}{l}{$\star$: GPT: GPT5-mini; Gemini: Gemini-2.0; DeepSeek: DeepSeek-V3.1. We set the CSA top-$k$ parameter to $k=4$.}

\end{tabular}
\end{table*}
% In this section, our experimental evaluation 
We investigate the following \textbf{research questions (RQs)}:
(RQ1) How does \sysname compare with other agent frameworks?
(RQ2) How effective are the three featured agents?
(RQ3) How does multi-level, multi-modal representation compare with flat, text-only input?
(RQ4) How does the top-$k$ factor affect context selection?
(RQ5) How do iteration limits and cost vary across tasks?
% \end{itemize}
\subsection{Overall Performance Comparison (RQ1)}
We evaluate our system on two benchmark levels: L1 (Table~\ref{tab:l1-eval}) and L2 (Table~\ref{tab:main-r2}). L1 focuses on fundamental agent abilities such as understanding design rules, interpreting geometric context, and generating valid tool calls. Large-scale models (GPT5-mini, Gemini-2.0, DeepSeek-V3.1) perform reliably, reaching overall pass@5 scores of \textbf{99.2\%}, \textbf{95.1\%}, and \textbf{94.2\%}, while smaller models (Qwen-2.5-7B, Llama-3.1-8B) frequently fail due to malformed or semantically incorrect tool invocations. Therefore, only the large-scale models are used for L2 evaluation.

L2 targets more challenging multi-step DRC-fixing tasks. \textbf{The errors at L2 are essentially the remaining violations that OpenROAD cannot fix automatically.} Our multi-agent system with GPT5-mini achieves the best performance, obtaining \textbf{30.4\%} pass@1 and \textbf{55.7\%} pass@5. For Class~1 tasks, ReAct(GPT5-mini) performs competitively (24.0\% pass@1 and 51.7\% pass@5) due to GPT5-mini’s strong intrinsic reasoning ability. As task complexity increases in Class~2, our algorithms consistently outperform single-agent methods across all base models---\sysname(GPT) achieving the best results and \sysname(DeepSeek) ranking second---demonstrating the benefits of coordinated multi-agent reasoning. For Class~3, the Gemini-based variant performs better (33.3\% pass@1 and 82.4\% pass@5), indicating Gemini's stronger exploratory behavior when the agent must reason beyond the given error description. However, for the planner–actor architecture, performance degrades on the three strong base models—GPT5-mini, Gemini-2.0, and DeepSeek-V3.1. These models already integrate reasoning and tool invocation effectively, so separating planning and execution introduces misalignment and additional errors. Overall, the results show that multi-agent collaboration of \sysname significantly improves the reliability of DRC fixing under complex scenarios.

\subsection{Ablation Study (RQ2)}
% To evaluate the contribution of each component in our multi-agent design, 
We conduct an ablation study using Gemini-2.0 as the base model by selectively removing the Observation Agent, Context Selection Agent, and Spatial-Awareness Agent. The experimental variants are listed below, and the results are reported in Table~\ref{tab:ablation_study}.

\begin{itemize}[leftmargin=*]
    \item \textbf{Baseline (ReAct).} A single-agent ReAct system that performs reasoning, tool invocation, and feedback interpretation without any specialized agent decomposition.
    \item \textbf{\sysname~w/o Observation Agent.} The system removes the module responsible for proposing the initial plan and candidate tool actions. Tool outputs are directly forwarded to the Context Selection Agent for shape filtering.
    \item \textbf{\sysname~w/o Context Selection Agent.} Instead of selecting relevant shapes, all geometric entities are passed to the Spatial-Awareness Agent, increasing contextual redundancy and forcing downstream agents to process unnecessary information.
    \item \textbf{\sysname~w/o Spatial-Awareness Agent.} Raw geometric data is delivered directly to the Checker Agent without structured spatial relationships among selected shapes.
    \item \textbf{\sysname.} The complete multi-agent system with all three featured agents.
\end{itemize}
\begin{table}[]
\caption{Ablation study on L2. Pass@5~$\Delta$ indicates the drop relative to the full \sysname model.}
\label{tab:ablation_study}
\begin{tabular}{lccc}
\hline
\multicolumn{1}{c}{\multirow{2}{*}{\textbf{Variants}}} & \multicolumn{2}{c}{\textbf{L2}} & \multirow{2}{*}{\textbf{Pass@5 $\Delta$}} \\ \cline{2-3}
\multicolumn{1}{c}{}                                   & pass@1         & pass@5         &                                   \\ \hline
Baseline (ReAct)                                       & 10.0           & 32.0           & $\downarrow$ 9.3                                  \\
w/o Observation Agent                                  & 18.3           & 38.7           & $\downarrow$ 2.6                                 \\
w/o Context Selection Agent                            & 11.3           & 30.9           & $\downarrow$ 10.4                                \\
w/o Spatial-Awareness Agent                            & 11.7           & 33.5           & $\downarrow$ 7.8                               \\
\sysname                                              & 14.8           & 41.3           &   $=$                               \\ \hline
\end{tabular}
\end{table}
As shown in Table~\ref{tab:ablation_study}, removing the OA results in only a moderate performance drop (18.3\% pass@1 and 38.7\% pass@5). This suggests that even without explicit initial planning, the downstream agents can still infer reasonable actions when sufficient spatial cues are retained. In contrast, omitting the Context Selection Agent causes the most significant degradation (10.4\% pass@5), demonstrating that filtering irrelevant shapes is critical for reducing contextual noise and enabling precise reasoning. Removing the Spatial-Awareness Agent also leads to noticeable degradation (7.8\% pass@5), confirming the importance of structured spatial relationships in guiding the checker toward accurate action proposals.

Overall, the complete \sysname configuration achieves the strongest results, validating the necessity and complementarity of all three featured agents within our multi-agent architecture.

\subsection{Effectiveness of Multi-Modality (RQ3)}
\begin{figure}[t]
\centering
\includegraphics[width=\linewidth]{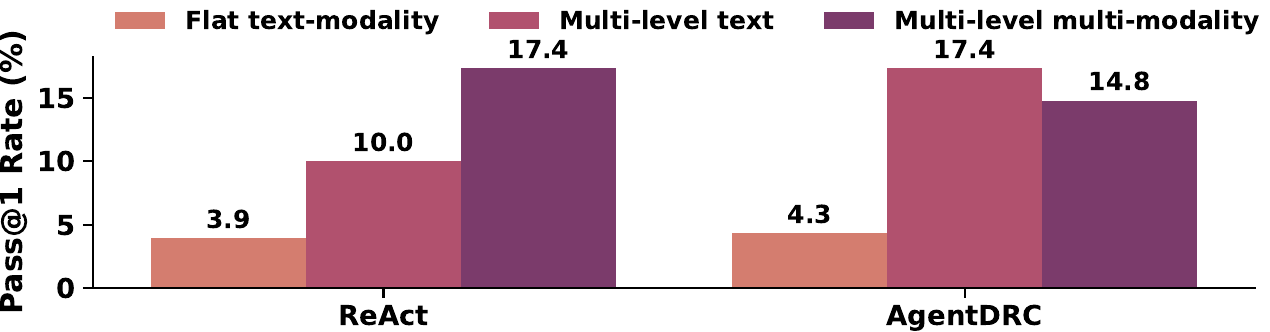} \\ \vspace{.5em}
\includegraphics[width=\linewidth]{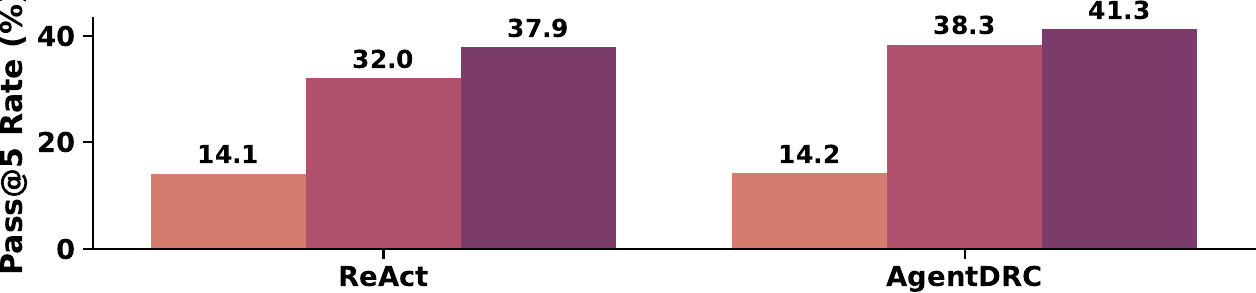}
\caption{\textbf{Effectiveness of Multi-Level and Multi-Modal Geometric Information}}
\label{fig:ablation_of_techniques}
\end{figure}

To evaluate the impact of geometric information structure, we compare flat text-modality, multi-level text, and multi-level multi-modal (text+vision) inputs under ReAct and \sysname using Gemini-2.0 (Fig.~\ref{fig:ablation_of_techniques}). With flat text input, both systems perform poorly (ReAct: 3.9\%/14.1\%, \sysname: 4.3\%/14.2\% on pass@1/pass@5), showing that the agent cannot reliably locate key shapes from flat raw coordinates. Introducing multi-level hierarchical text substantially improves performance, raising ReAct to 10.0\%/32.0\% and \sysname to 17.4\%/38.3\%, as structured multi-level geometric information help the agent focus on relevant geometry. 

Adding multi-modal vision yields further gains (ReAct: 17.4\%/37.9\%, \sysname: 14.8\%/41.3\%), although the improvement is more pronounced for ReAct because \sysname already benefits from its Context Selection and Spatial-Awareness agents. Overall, multi-level multi-modality provides the strongest results, confirming that hierarchical and visual cues jointly enhance spatial reasoning.

\begin{figure}[t]
\centering
\includegraphics[width=\linewidth]{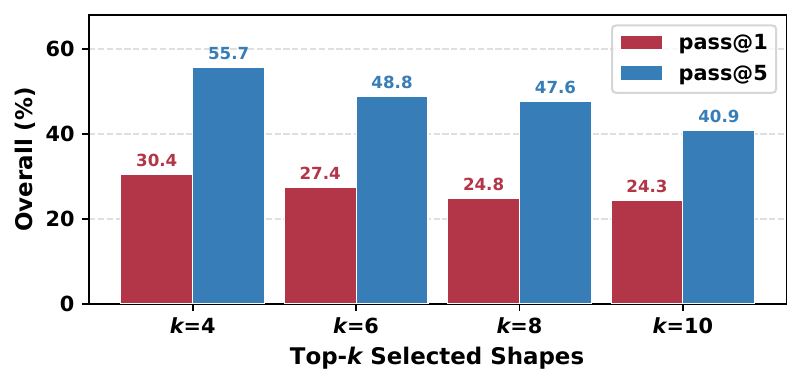}
\caption{\textbf{Effect of top-$k$ selected shapes on L2 overall performance.} Fewer selected shapes yield better results, as excessive context introduces noise.}
\label{fig:rq4_topk}
\end{figure}

\subsection{Effect of Top-$k$ Selection (RQ4)}
To study the sensitivity of the CSA's top-$k$ parameter, we vary the number of selected neighbor context shapes from $k{=}4$ to $k{=}10$ while keeping all other settings fixed (\sysname with GPT5-mini, on L2). Results are shown in Fig.~\ref{fig:rq4_topk}.

With $k{=}4$, the system achieves the best overall performance (30.4\% pass@1, 55.7\% pass@5). Increasing $k$ to 6 and 8 leads to gradual degradation, and at $k{=}10$ performance drops notably (24.3\% pass@1, 40.9\% pass@5). This trend confirms that including more shapes introduces contextual noise: while a small number of carefully selected shapes provides the checker with focused and relevant spatial information, a larger context dilutes the critical cues and impairs reasoning accuracy. Therefore, we adopt $k{=}4$ as the default throughout our experiments, achieving the optimal trade-off between coverage and precision.

\subsection{Effect of Max Iterations (RQ5)}
\begin{figure}[t]
\centering
\includegraphics[width=\linewidth]{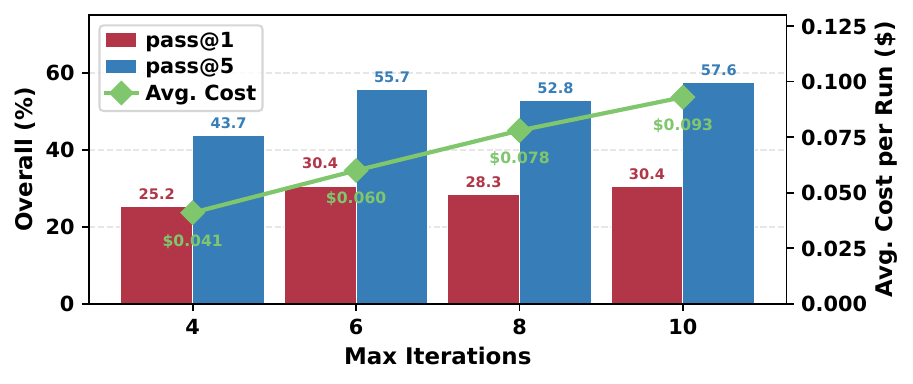}
\caption{\textbf{Effect of max iterations on L2 performance and cost.} (a) Overall pass@1 and pass@5 largely saturate beyond 6 iterations. (b) API cost per run increases steadily with the iteration budget.}
\label{fig:rq5_iterations}
\end{figure}

We vary the iteration limit from 4 to 10 under \sysname with GPT5-mini (Fig.~\ref{fig:rq5_iterations}), where one iteration is a complete pass through the full agent pipeline---observation, context selection, spatial-aware verification, and tool execution.

With only 4 iterations, performance is the lowest (25.2\% pass@1, 43.7\% pass@5), as the agent often exhausts its budget before completing multi-step fixes. Increasing the limit to 6 yields a substantial jump (30.4\% pass@1, 55.7\% pass@5). Beyond 6 iterations, the returns diminish: we observe that tasks whose initial reasoning direction is correct tend to succeed within 6 iterations, while tasks that start on a wrong path rarely recover even with additional iterations. Specifically, 8 iterations slightly underperforms 6 (28.3\% pass@1, 52.8\% pass@5) due to over-exploration, while 10 iterations recovers to 30.4\% pass@1 and 57.6\% pass@5.

The API cost grows roughly linearly with the iteration cap, ranging from \$0.041 at 4 iterations to \$0.093 at 10 iterations, which remains affordable across all settings. We adopt 6 as the default, offering the best performance--cost trade-off.

%% file: Sections/06_conclusion.tex
\section{Conclusion}
We present \sysname, a set of algorithms for spatially-aware DRC correction that coordinate multiple LLM agents to fix violations unresolved by conventional EDA tools. The system perceives multi-level, multi-modal geometric context, filters irrelevant shapes via two-step context selection, reconstructs structured spatial relations among critical shapes, and verifies proposed actions through geometric and semantic constraint checking. Experiments show that \sysname achieves up to 30.4\% pass@1 and 55.7\% pass@5 on challenging violations, confirming the effectiveness of both the proposed algorithms and the multi-agent coordination.

%% file: reference.bib
@inproceedings{
hong2024metagpt,
title={Meta{GPT}: Meta Programming for A Multi-Agent Collaborative Framework},
author={Sirui Hong and Mingchen Zhuge and Jonathan Chen and Xiawu Zheng and Yuheng Cheng and Jinlin Wang and Ceyao Zhang and Zili Wang and Steven Ka Shing Yau and Zijuan Lin and Liyang Zhou and Chenyu Ran and Lingfeng Xiao and Chenglin Wu and J{\"u}rgen Schmidhuber},
booktitle={The Twelfth International Conference on Learning Representations},
year={2024},
url={https://openreview.net/forum?id=VtmBAGCN7o}
}

@inproceedings{zhang-etal-2024-codeagent,
    title = "{C}ode{A}gent: Enhancing Code Generation with Tool-Integrated Agent Systems for Real-World Repo-level Coding Challenges",
    author = "Zhang, Kechi  and
      Li, Jia  and
      Li, Ge  and
      Shi, Xianjie  and
      Jin, Zhi",
    editor = "Ku, Lun-Wei  and
      Martins, Andre  and
      Srikumar, Vivek",
    booktitle = "Proceedings of the 62nd Annual Meeting of the Association for Computational Linguistics (Volume 1: Long Papers)",
    month = aug,
    year = "2024",
    address = "Bangkok, Thailand",
    publisher = "Association for Computational Linguistics",
    url = "https://aclanthology.org/2024.acl-long.737/",
    doi = "10.18653/v1/2024.acl-long.737",
    pages = "13643--13658"
}

@inproceedings{
chen2024teaching,
title={Teaching Large Language Models to Self-Debug},
author={Xinyun Chen and Maxwell Lin and Nathanael Sch{\"a}rli and Denny Zhou},
booktitle={The Twelfth International Conference on Learning Representations},
year={2024},
url={https://openreview.net/forum?id=KuPixIqPiq}
}

@inproceedings{yao2023react,
  title = {{ReAct}: Synergizing Reasoning and Acting in Language Models},
  author = {Yao, Shunyu and Zhao, Jeffrey and Yu, Dian and Du, Nan and Shafran, Izhak and Narasimhan, Karthik and Cao, Yuan},
  booktitle = {International Conference on Learning Representations (ICLR) },
  year = {2023},
  html = {https://arxiv.org/abs/2210.03629},
}

@inproceedings{palm,
author = {Driess, Danny and Xia, Fei and Sajjadi, Mehdi S. M. and Lynch, Corey and Chowdhery, Aakanksha and Ichter, Brian and Wahid, Ayzaan and Tompson, Jonathan and Vuong, Quan and Yu, Tianhe and Huang, Wenlong and Chebotar, Yevgen and Sermanet, Pierre and Duckworth, Daniel and Levine, Sergey and Vanhoucke, Vincent and Hausman, Karol and Toussaint, Marc and Greff, Klaus and Zeng, Andy and Mordatch, Igor and Florence, Pete},
title = {PaLM-E: an embodied multimodal language model},
year = {2023},
publisher = {JMLR.org},
booktitle = {Proceedings of the 40th International Conference on Machine Learning},
articleno = {340},
numpages = {20},
location = {Honolulu, Hawaii, USA},
series = {ICML'23}
}

@inproceedings{toolformer,
author = {Schick, Timo and Dwivedi-Yu, Jane and Dess\'{\i}, Roberto and Raileanu, Roberta and Lomeli, Maria and Hambro, Eric and Zettlemoyer, Luke and Cancedda, Nicola and Scialom, Thomas},
title = {Toolformer: language models can teach themselves to use tools},
year = {2023},
publisher = {Curran Associates Inc.},
address = {Red Hook, NY, USA},
booktitle = {Proceedings of the 37th International Conference on Neural Information Processing Systems},
articleno = {2997},
numpages = {13},
location = {New Orleans, LA, USA},
series = {NIPS '23}
}

@inproceedings{pal,
author = {Gao, Luyu and Madaan, Aman and Zhou, Shuyan and Alon, Uri and Liu, Pengfei and Yang, Yiming and Callan, Jamie and Neubig, Graham},
title = {PAL: program-aided language models},
year = {2023},
publisher = {JMLR.org},
booktitle = {Proceedings of the 40th International Conference on Machine Learning},
articleno = {435},
numpages = {36},
location = {Honolulu, Hawaii, USA},
series = {ICML'23}
}

@inproceedings{
patil2024gorilla,
title={Gorilla: Large Language Model Connected with Massive {API}s},
author={Shishir G Patil and Tianjun Zhang and Xin Wang and Joseph E. Gonzalez},
booktitle={The Thirty-eighth Annual Conference on Neural Information Processing Systems},
year={2024},
url={https://openreview.net/forum?id=tBRNC6YemY}
}

@techreport{siemens2020drc,
  author       = {Siemens Digital Industries Software},
  title        = {Automated post-processing of DRC errors improves debugging productivity},
  year         = {2020},
  institution  = {Siemens EDA},
  note         = {White Paper},
  url          = {https://static.sw.cdn.siemens.com/siemens-disw-assets/public/5R5kzgoX1k13WtoqjEEFqO/en-US/Siemens-SW-Automated-post-processing-DRC-debugging-83714.pdf}
}

@techreport{gfmentor2019drcfix,
  author       = {GlobalFoundries and Mentor, A Siemens Business},
  title        = {Automated DRC Violation Correction Speeds Designs to Tapeout},
  year         = {2019},
  note         = {White Paper},
  url          = {https://www.edn.com/automated-drc-violation-correction-speeds-designs-to-tapeout/}
}

@techreport{siemens2018pdk,
  author       = {Siemens Digital Industries Software},
  title        = {The Process Design Kit: Protecting Design Know-How},
  year         = {2018},
  institution  = {Siemens EDA},
  note         = {White paper},
  url          = {https://static.sw.cdn.siemens.com/siemens-disw-assets/public/82831/en-US/Siemens-SW-The-process-design-kit-WP-81843-C2.pdf}
}

@article{clark2016asap7,
  author       = {Clark, L.,T. and Vashishtha, V. and Shifren, L. and Gujja, A. and Sinha, S. and Cline, B. and Ramamurthy, C. and Yeric, G.},
  title        = {ASAP7: A 7-nm finFET predictive process design kit},
  journal      = {Microelectronics Journal},
  volume       = {53},
  number       = {—},
  pages        = {105--115},
  year         = {2016},
  doi          = {10.1016/j.mejo.2016.04.006}
}

@techreport{cadence2023innovus,
  author       = {Cadence Design Systems, Inc.},
  title        = {Cadence Innovus+ Synthesis and Implementation System - Integrated RTL to GDS Chip Design Delivering Accelerated Results},
  year         = {2023},
  institution  = {Cadence Design Systems, Inc.},
  note         = {White Paper},
  url          = {https://www.cadence.com/en_US/home/resources/white-papers/innovus-plus-synthesis-implementation-system-wp.html}
}

@techreport{mentor2016calibre,
  author       = {Mentor Graphics, A Siemens Business},
  title        = {Calibre Physical Verification System: Design Rule Checking and Signoff Verification},
  year         = {2016},
  institution  = {Mentor Graphics Corporation},
  note         = {Technical White Paper},
  url          = {https://resources.sw.siemens.com/en-US/white-paper-calibre-physical-verification-system}
}

@inproceedings{openroad2020,
  author       = {Ajayi, T. and Kahng, A. B. and Li, B. and Nath, R. and Sharma, P. and Soroushmehr, M. and Xu, B.},
  title        = {OpenROAD: Toward a Self-Driving, Open-Source Digital Layout Implementation Tool Chain},
  booktitle    = {Proceedings of the ACM/IEEE Design Automation Conference (DAC)},
  year         = {2020},
  pages        = {1--6},
  doi          = {10.1109/DAC18072.2020.9218696}
}

@article{zhang2021multiagent,
  author    = {Zhang, Kaiqing and Yang, Zhuoran and Basar, Tamer},
  title     = {Multi-Agent Reinforcement Learning: A Selective Overview of Theories and Algorithms},
  journal   = {Handbook of Reinforcement Learning and Control},
  pages     = {321--384},
  year      = {2021},
  publisher = {Springer}
}

@inproceedings{bo2024reflective,
  author    = {Bo, Xiaohe and Zhang, Zeyu and Dai, Quanyu and Feng, Xueyang and Wang, Lei and Li, Rui and Chen, Xu and Wen, Ji-Rong},
  title     = {Reflective Multi-Agent Collaboration based on Large Language Models},
  booktitle = {Proceedings of the 38th Conference on Neural Information Processing Systems (NeurIPS)},
  year      = {2024},
}

@inproceedings{nayak2024llamar,
  author    = {Nayak, S. and others},
  title     = {Long-Horizon Planning for Multi-Agent Robots in Partially Observable Environments},
  booktitle = {Proceedings of the 38th Conference on Neural Information Processing Systems (NeurIPS)},
  year      = {2024},
}

@inproceedings{wang2024voyager,
  author    = {Wang, Guanzhi and Fan, Jiahui and Qian, Chen and Liu, Weiyu and Park, Jiwoo and Zhou, Yichi and Liu, Yuxuan and others},
  title     = {Voyager: An Open-Ended Embodied Agent with Large Language Models},
  booktitle = {Proceedings of the 41st International Conference on Machine Learning (ICML)},
  year      = {2024},
  note      = {Employs LLM-based planner–executor framework for open-world task completion},
  url       = {https://arxiv.org/abs/2305.16291}
}

@InProceedings{song2023llmplanner,
  author    = {Song, Chan Hee and Wu, Jiaman and Washington, Clayton and Sadler, Brian M. and Chao, Wei-Lun and Su, Yu},
  title     = {LLM-Planner: Few-Shot Grounded Planning for Embodied Agents with Large Language Models},
  booktitle = {Proceedings of the IEEE/CVF International Conference on Computer Vision (ICCV)},
  month     = {October},
  year      = {2023},
}

@article{chen2021codex,
  title={Evaluating Large Language Models Trained on Code},
  author={Mark Chen and Jerry Tworek and Heewoo Jun and Qiming Yuan and Henrique Ponde de Oliveira Pinto and Jared Kaplan and Harri Edwards and Yuri Burda and Nicholas Joseph and Greg Brockman and Alex Ray and Raul Puri and Gretchen Krueger and Michael Petrov and Heidy Khlaaf and Girish Sastry and Pamela Mishkin and Brooke Chan and Scott Gray and Nick Ryder and Mikhail Pavlov and Alethea Power and Lukasz Kaiser and Mohammad Bavarian and Clemens Winter and Philippe Tillet and Felipe Petroski Such and Dave Cummings and Matthias Plappert and Fotios Chantzis and Elizabeth Barnes and Ariel Herbert-Voss and William Hebgen Guss and Alex Nichol and Alex Paino and Nikolas Tezak and Jie Tang and Igor Babuschkin and Suchir Balaji and Shantanu Jain and William Saunders and Christopher Hesse and Andrew N. Carr and Jan Leike and Josh Achiam and Vedant Misra and Evan Morikawa and Alec Radford and Matthew Knight and Miles Brundage and Mira Murati and Katie Mayer and Peter Welinder and Bob McGrew and Dario Amodei and Sam McCandlish and Ilya Sutskever and Wojciech Zaremba},
  year={2021},
  eprint={2107.03374},
  archivePrefix={arXiv},
  primaryClass={cs.LG}
}

@techreport{siemens2020sadp,
  title        = {Self-aligned double patterning (SADP) design finishing: Improving results with DRC auto-fixing},
  author       = {{GLOBALFOUNDRIES and Siemens EDA}},
  institution  = {Siemens Digital Industries Software},
  year         = {2020},
  url          = {https://resources.sw.siemens.com/en-US/technical-paper-self-aligned-double-patterning-design-finishing-improving-results-with-drc},
  note         = {Technical paper describing challenges and automated DRC fixing limitations in SADP layouts}
}

@techreport{siemens2023drcdebug,
  title        = {Automated Post-Processing for DRC Debugging},
  author       = {{Siemens EDA}},
  institution  = {Siemens Digital Industries Software},
  year         = {2023},
  url          = {https://static.sw.cdn.siemens.com/siemens-disw-assets/public/5R5kzgoX1k13WtoqjEEFqO/en-US/Siemens-SW-Automated-post-processing-DRC-debugging-83714.pdf},
  note         = {White paper explaining DRC/DFM complexity and the impracticality of achieving fully clean designs purely through automation}
}

@misc{klayout2024,
  author       = {Matthias K\"{o}fferlein},
  title        = {{KLayout} -- High Performance Layout Viewer and Editor},
  year         = {2024},
  url          = {https://www.klayout.de/},
  note         = {Open-source layout viewer and editor}
}
